\documentclass[twocolumn,prd,superscriptaddress,altaffilletter,nofootinbib]{revtex4-1}

\usepackage{amsmath}
\usepackage{amssymb}
\usepackage{amsfonts}
\usepackage{graphicx}
\usepackage{hyperref}
\usepackage{lineno}
\usepackage[usenames,dvipsnames]{xcolor}
\usepackage{xspace}
\usepackage{tikz}
\usetikzlibrary{arrows,shapes,trees,decorations.pathreplacing}

\allowdisplaybreaks

\allowdisplaybreaks

\newcommand{\treebank}{\texttt{treebank}}

\newcommand{\g}{\ensuremath{\mathbf{g}}}

\newcommand{\x}{\ensuremath{\vec{\lambda}}}

\newcommand{\dx}{\ensuremath{\vec{\Delta \lambda}}}

\newcommand{\pdim}{\ensuremath{D}}
\newcommand{\bifur}{\ensuremath{n}}
\newcommand{\numper}[1]{\ensuremath{\mathcal{N}_{#1}}}

\begin{document}

\title{A binary tree approach to template placement for searches for
  gravitational waves from compact binary mergers}

\author{Chad Hanna}
\affiliation{Department of Physics, The Pennsylvania State University, University Park, PA 16802, USA}
\affiliation{Institute for Gravitation and the Cosmos, The Pennsylvania State University, University Park, PA 16802, USA}
\affiliation{Department of Astronomy and Astrophysics, The Pennsylvania State University, University Park, PA 16802, USA}
\affiliation{Institute for Computational and Data Sciences, The Pennsylvania State University, University Park, PA 16802, USA}

\author{James Kennington}
\email{james.kennington@ligo.org}
\affiliation{Department of Physics, The Pennsylvania State University, University Park, PA 16802, USA}
\affiliation{Institute for Gravitation and the Cosmos, The Pennsylvania State University, University Park, PA 16802, USA}

\author{Shio Sakon}
\affiliation{Department of Physics, The Pennsylvania State University, University Park, PA 16802, USA}
\affiliation{Institute for Gravitation and the Cosmos, The Pennsylvania State University, University Park, PA 16802, USA}

\author{Stephen Privitera}
\email{stephen.privitera@ligo.org}
\affiliation{Albert-Einstein-Institut, Max-Planck-Institut f{\"u}r Gravitationsphysik, D-14476 Potsdam-Golm, Germany}

\author{Miguel Fernandez}
\affiliation{Department of Physics, The Pennsylvania State University, University Park, PA 16802, USA}
\affiliation{Institute for Gravitation and the Cosmos, The Pennsylvania State University, University Park, PA 16802, USA}

\author{Jonathan Wang}
\affiliation{Department of Physics, University of Michigan, Ann Arbor, Michigan 48109, USA}

\author{Cody Messick}
\affiliation{Department of Physics, The Pennsylvania State University, University Park, PA 16802, USA}

\author{Alex Pace}
\affiliation{Department of Physics, The Pennsylvania State University, University Park, PA 16802, USA}
\affiliation{Institute for Gravitation and the Cosmos, The Pennsylvania State University, University Park, PA 16802, USA}

\author{Kipp Cannon}
\affiliation{RESCEU, The University of Tokyo, Tokyo, 113-0033, Japan}

\author{Prathamesh Joshi}
\affiliation{Department of Physics, The Pennsylvania State University, University Park, PA 16802, USA}
\affiliation{Institute for Gravitation and the Cosmos, The Pennsylvania State University, University Park, PA 16802, USA}

\author{Rachael Huxford}
\affiliation{Department of Physics, The Pennsylvania State University, University Park, PA 16802, USA}
\affiliation{Institute for Gravitation and the Cosmos, The Pennsylvania State University, University Park, PA 16802, USA}

\author{Sarah Caudill}
\affiliation{Nikhef, Science Park, 1098 XG Amsterdam, Netherlands}

\author{Chiwai Chan}
\affiliation{RESCEU, The University of Tokyo, Tokyo, 113-0033, Japan}

\author{Bryce Cousins}
\affiliation{Department of Physics, The Pennsylvania State University, University Park, PA 16802, USA}
\affiliation{Institute for Computational and Data Sciences, The Pennsylvania State University, University Park, PA 16802, USA}

\author{Jolien D. E. Creighton}
\affiliation{Leonard E.\ Parker Center for Gravitation, Cosmology, and Astrophysics, University of Wisconsin-Milwaukee, Milwaukee, WI 53201, USA}

\author{Becca Ewing}
\affiliation{Department of Physics, The Pennsylvania State University, University Park, PA 16802, USA}
\affiliation{Institute for Gravitation and the Cosmos, The Pennsylvania State University, University Park, PA 16802, USA}

\author{Heather Fong}
\affiliation{RESCEU, The University of Tokyo, Tokyo, 113-0033, Japan}
\affiliation{Graduate School of Science, The University of Tokyo, Tokyo 113-0033, Japan}

\author{Patrick Godwin}
\affiliation{Department of Physics, The Pennsylvania State University, University Park, PA 16802, USA}
\affiliation{Institute for Gravitation and the Cosmos, The Pennsylvania State University, University Park, PA 16802, USA}

\author{Ryan Magee}
\affiliation{Department of Physics, The Pennsylvania State University, University Park, PA 16802, USA}
\affiliation{Institute for Gravitation and the Cosmos, The Pennsylvania State University, University Park, PA 16802, USA}

\author{Duncan Meacher}
\affiliation{Leonard E.\ Parker Center for Gravitation, Cosmology, and Astrophysics, University of Wisconsin-Milwaukee, Milwaukee, WI 53201, USA}

\author{Soichiro Morisaki}
\affiliation{Institute for Cosmic Ray Research, The University of Tokyo, 5-1-5 Kashiwanoha, Kashiwa, Chiba 277-8582, Japan}

\author{Debnandini Mukherjee}
\affiliation{Department of Physics, The Pennsylvania State University, University Park, PA 16802, USA}
\affiliation{Institute for Gravitation and the Cosmos, The Pennsylvania State University, University Park, PA 16802, USA}

\author{Hiroaki Ohta}
\affiliation{RESCEU, The University of Tokyo, Tokyo, 113-0033, Japan}

\author{Surabhi Sachdev}
\affiliation{Department of Physics, The Pennsylvania State University, University Park, PA 16802, USA}
\affiliation{Institute for Gravitation and the Cosmos, The Pennsylvania State University, University Park, PA 16802, USA}
\affiliation{LIGO Laboratory, California Institute of Technology, MS 100-36, Pasadena, California 91125, USA}

\author{Divya Singh}
\affiliation{Department of Physics, The Pennsylvania State University, University Park, PA 16802, USA}
\affiliation{Institute for Gravitation and the Cosmos, The Pennsylvania State University, University Park, PA 16802, USA}

\author{Ron Tapia}
\affiliation{Department of Physics, The Pennsylvania State University, University Park, PA 16802, USA}
\affiliation{Institute for Computational and Data Sciences, The Pennsylvania State University, University Park, PA 16802, USA}

\author{Leo Tsukada}
\affiliation{RESCEU, The University of Tokyo, Tokyo, 113-0033, Japan}
\affiliation{Graduate School of Science, The University of Tokyo, Tokyo 113-0033, Japan}

\author{Daichi Tsuna}
\affiliation{RESCEU, The University of Tokyo, Tokyo, 113-0033, Japan}
\affiliation{Graduate School of Science, The University of Tokyo, Tokyo 113-0033, Japan}

\author{Takuya Tsutsui}
\affiliation{RESCEU, The University of Tokyo, Tokyo, 113-0033, Japan}

\author{Koh Ueno}
\affiliation{RESCEU, The University of Tokyo, Tokyo, 113-0033, Japan}

\author{Aaron Viets}
\affiliation{Leonard E.\ Parker Center for Gravitation, Cosmology, and Astrophysics, University of Wisconsin-Milwaukee, Milwaukee, WI 53201, USA}

\author{Leslie Wade}
\affiliation{Department of Physics, Hayes Hall, Kenyon College, Gambier, Ohio 43022, USA}

\author{Madeline Wade}
\affiliation{Department of Physics, Hayes Hall, Kenyon College, Gambier, Ohio 43022, USA}


\date{\today}

\begin{abstract}
We demonstrate a new geometric method for fast template placement for searches
for gravitational waves from the inspiral, merger and ringdown of compact
binaries. The method is based on a binary tree decomposition of the template
bank parameter space into non-overlapping hypercubes. We use a numerical
approximation of the signal overlap metric at the center of each hypercube to
estimate the number of templates required to cover the hypercube and determine
whether to further split the hypercube. As long as the expected number of
templates in a given cube is greater than a given threshold, we split the cube
along its longest edge according to the metric. When the expected number of
templates in a given hypercube drops below this threshold, the splitting stops
and a template is placed at the center of the hypercube. Using this method, we
generate aligned-spin template banks covering the mass range suitable for a
search of Advanced LIGO data. The aligned-spin bank required $\sim 24$
CPU-hours and produced 2 million templates.  In general, we find that other
methods, namely stochastic placement, produces a more strictly bounded loss in
match between waveforms, with the same minimal match between waveforms
requiring about twice as many templates with our proposed algorithm. Though we
note that the average match is higher, which would lead to a higher detection
efficiency.  Our primary motivation is not to strictly minimize the number of
templates with this algorithm, but rather to produce a bank with useful
geometric properties in the physical parameter space coordinates. Such
properties are useful for population modeling and parameter estimation.
\end{abstract}

\maketitle

\section{Introduction}
Banks of template gravitational-wave signals are central tools in the
matched-filter detection of gravitational-wave signals from compact binary
coalescence~\cite{Sathyaprakash:1991mt, Owen:1995tm, Owen:1998dk}. The general
compact binary gravitational-wave signal depends on at least fifteen
parameters: two mass parameters, six spin parameters, distance, time, and five
angles defining binary orientation with respect to the gravitational-wave
antenna. The parameter space can be even larger if, for instance, matter or
eccentricity effects are included. Since we do not know the source parameters
{\it a priori}, we must search the data over all possible source parameters.

We are often able to quickly maximize the signal-to-noise ratio (SNR) over a
subset of the parameters either analytically or by efficient numerical
techniques. For instance, some parameters\footnote{Which parameters these are
depends on the assumptions made about the signal. For instance, non-precessing
binaries have a constant inclination angle, which enters into the
gravitational-wave signal only in the overall scale of the waveform, whereas
precessing binaries have a time-dependent inclination, leading to modulation in
the waveform phase and amplitude.} enter only into the overall amplitude of the
signal which is normalized away by the matched-filter definition of SNR.  The
coalescence time enters into the waveform as a frequency-dependent phase shift
which can efficiently be searched over using widely-available fast Fourier
Transform routines.  Considering only dominant $(\ell, |m|) = (2,2)$ modes of
gravitational-wave signals, the coalescence phase can also be maximized over
analytically.

Given the approximations, assumptions and techniques described above, a subset
of parameters, $\x{}$, the {\it template bank} parameters, are generally
relevant for template placement. We search over these parameters by laying down
a discrete set of points in the parameter space and repeating the
matched-filter calculation for each template. The set of points must be chosen
as a compromise between optimal SNR recovery and available computational
resources. Placing templates finely in the template parameter space leads to
high SNR recovery, but can quickly make the search prohibitively expensive. In
particular, the number of templates required to cover an $\pdim$-dimensional
parameter space such that no more than a fraction $M$ of the SNR is lost to any
potential signal scales as $M^{-\pdim/2}$~\cite{Owen:1995tm}.
   
In the case of non-spinning binaries, lattice placement strategies based on an
approximate analytic expression for the signal space ``distance'' between two
nearby templates have been shown to be effective for covering the template
parameter space~\cite{Cokelaer:2007kx, Abbott:2007ai}. To guarantee efficiency
of the placement, these methods require that the metric $\g(\x)$, which defines
the distance between nearby templates, is very nearly constant throughout the
parameter space. For waveforms involving spin, in which a metric is either
unavailable or varies rapidly throughout the parameter space, stochastic
template placement has proven to be effective in covering the parameter
space~\cite{Harry:2009ea, Babak:2008rb, Manca:2009xw, Ajith:2012mn,
Privitera:2013xza}. The stochastic placement technique works by randomly
selecting a large number of points in parameter space and keeping only those
points which fall sufficiently far away from points which have already been
accepted into the bank. This technique, while robust, is computationally
inefficient, although recent implementations have made significant strides
towards optimization~\cite{Ajith:2012mn, Capano:2016dsf, Fehrmann:2014cpa}.


Geometric techniques have also been applied to generate aligned-spin template
banks~\cite{Harry:2013tca, Roy:2017qgg}. In Ref.~\cite{Harry:2013tca}, the
authors demonstrate a geometric template bank for neutron-star--black-hole
binaries. The authors find satisfactory coverage for this parameter space by
stacking two two-dimensional lattices, taking advantage of the fact that the
parameter space is ``thin'' in the third dimension. This placement strategy was
used in conjunction with ordinary stochastic placement~\cite{Capano:2016dsf} to
cover the full compact binary parameter space searched in the recent LIGO-Virgo
searches~\cite{TheLIGOScientific:2016pea, Abbott:2016ymx}. In
Ref.~\cite{Roy:2017qgg}, the authors consider an interesting extension of this
technique which starts with a true three-dimensional lattice, and falls back to
the stochastic approach when the lattice approach breaks down. In
Ref.~\cite{Fehrmann:2014cpa}, the authors also consider a hybrid
stochastic-geometric technique, similar to the algorithm we propose here; however, the notion of lattice-adjacency the authors used is Cartesian whereas we incorporate the intrinsic geometry of the parameter manifold.

These solutions continue to rely at least partially on stochastic placement
methods, which scales poorly with the number of templates. The required number
of template parameters to cover a parameter space at a given minimal match
threshold increases dramatically with the bandwidth of the interferometer and
the dimension of the target signal space, both of which are ever-increasing in
ground-based gravitational wave searches~\cite{Harry:2010zz, Capano:2016dsf,
Harry:2016ijz}.  Currently used aligned-spin template banks have four template
parameters (two masses and two spins) and over 1 million templates at maximal
mismatches between 1--3\%~\cite{Mukherjee:2018yra}. Precessional effects adds five
more parameters (four spin components and the binary inclination at some
reference frequency) and an additional order of magnitude in
templates~\cite{Harry:2016ijz}.  At high mass ratios, sub-dominant modes may
also be important for detection, which can only further increase the template
bank size.  Presently template bank generation with stochastic methods may be
computationally slow. Future larger banks will require more computing resources
to generate as gravitational wave detector sensitivity improves.  This can be
problematic if banks are generated often.

Here, we demonstrate a new method for template placement based on a binary tree
decomposition of the parameter space which is purely geometric originally explored
here~\cite{wangthesis}. The algorithm
relies on a numerical estimation of the parameter space metric and uses this
metric to determine how to grow the binary tree. This algorithm requires
$\mathcal{O}(2^{\bifur} \pdim^2)$ overlap calculations, where $\bifur$ is the
bifurcation number of the parameter space, i.e., how many times a
characteristic cell is split, and $\dim$ is the dimension of the resulting
template bank. We demonstrate this method by constructing a bank suitable for
Advanced LIGO and Advanced Virgo data analysis.

\section{Motivation}

Beyond general interest in pursuing novel template placement algorithms, our
motivation for pursuing this work is three-fold based on experiences analyzing
LIGO and Virgo data during the third observing run.  First, in order to apply a
population model to gravitational wave detection, it is important to account
for template placement~\cite{Dent:2013cva} in a way that may account for the
coordinate volume that a template occupies~\cite{fong2018simulations, Magee:2019vmb,
LIGOScientific:2020ibl, LIGOScientific:2021djp}. The binary tree approach that
we have taken guarantees that each template ends up in a hyperrectangle in the
physical coordinates making coordinate volume calculations easy. Second, in
order to ensure a high availability of service for online compact binary
searches we run searches at two different data centers. The goal is to split
the parameter space in a way that if one site goes down the other is still
efficient at detecting a broad class of binary signals.  The binary tree
approach allows us to use a bank derived from the ``right" and ``left" splits
separately.  Finally, having a bank that is grid-like in physical coordinates
is generally useful for template interpolation~\cite{Cannon:2011rj} and rapid
parameter estimation~\cite{PhysRevD.92.023002} problems and we are interested
in exploring this as future work.

\section{Methods}

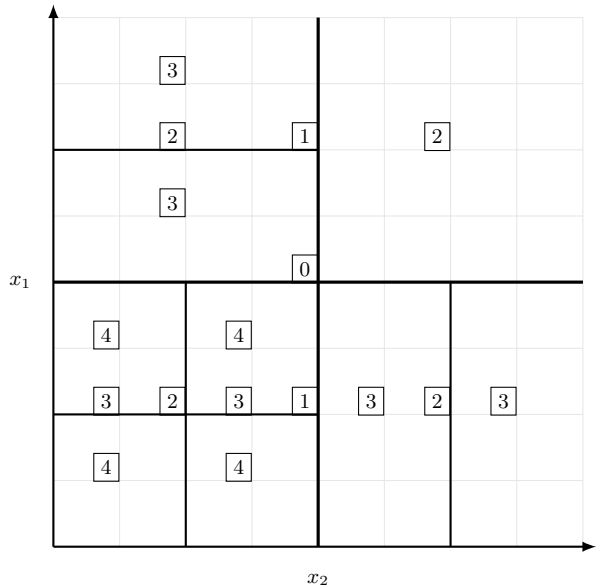
\begin{figure}
  \centering
  \begin{tikzpicture}[scale=0.88, transform shape, >=latex]
        \draw[step=1cm,gray!20,very thin] (0,0) grid (8,8);
        \draw[thick,->] (0,0) -- (8.2,0) node[anchor=north west] {};
        \draw[thick,->] (0,0) -- (0,8.2) node[anchor=south east] {};
	\node[] at (-0.5, 4.0) {$x_1$};
	\node[] at (4.0, -0.5) {$x_2$};
	\node[] at (-1.25, 4.0) {};
	\node[] at (4.0, -1.25) {};
        \draw[very thick] (0,4) -- (8,4) node[anchor=south east] {};
        \draw[very thick] (4,0) -- (4,8) node[anchor=south east] {};
        \draw[thick] (0,2) -- (4,2) node[anchor=south east] {};
        \draw[thick] (2,0) -- (2,4) node[anchor=south east] {};
        \draw[thick] (0,6) -- (4,6) node[anchor=south east] {};
        \draw[thick] (6,0) -- (6,4) node[anchor=south east] {};





	\node[draw] at (3.8,4.2) {0};

	\node[draw] at (3.8,6.2) {1};
	\node[draw] at (3.8,2.2) {1};

	\node[draw] at (1.8,6.2) {2};
	\node[draw, fill=white] at (5.8,6.2) {2};
	\node[draw] at (1.8,2.2) {2};
	\node[draw] at (5.8,2.2) {2};

	\node[draw] at (4.8,2.2) {3};
	\node[draw] at (6.8,2.2) {3};
	\node[draw] at (0.8,2.2) {3};
	\node[draw] at (2.8,2.2) {3};
	\node[draw] at (1.8,5.2) {3};
	\node[draw] at (1.8,7.2) {3};

	\node[draw] at (0.8,1.2) {4};
	\node[draw] at (0.8,3.2) {4};
	\node[draw] at (2.8,1.2) {4};
	\node[draw] at (2.8,3.2) {4};

\end{tikzpicture}
  \caption{\label{fig:method_illustration} Example hyper-rectangle
    bifurcation in two arbitrary dimensions, $x_1$, $x_2$.  Each number,
    $\fbox{n}$, represents a place where the metric, $\g$ was computed at the
    $nth$ stage of the bifurcation.  This example results in nine
    hyper-rectangles, which is less than the maximum value of $2^4$ after four
    bifurcations.  
    }
\end{figure}

Our method, whose implementation we refer to as \treebank, relies on having an
accurate approximation of the template space metric $\g(\x)$, which gives a
measure of the ``distance'' between nearby templates.  For our work $\x \equiv
\{t_c, \log m_1, \log m_2, \chi_\text{eff}\}$, where $\chi_\text{eff} \equiv (m_1
a_{1z} + m_2 a_{2z}) \, / \, (m_1 + m_2)$ and $a$ is the dimensionless
spin~\cite{ajithInspiralMergerRingdownWaveformsBlackHole2011}.  We define the mismatch $\delta^2$ between two nearby
gravitational-wave templates, $h(\x)$ and $h(\x + \dx)$, according to
\begin{align}
  \delta(\x, \dx)^2 &= 1 - \langle \, \hat{h}(\x) \, | \, \hat{h}(\x + \dx) \, \rangle, \\
  \langle a | b \rangle &\equiv \left| \int_{-f_N}^{f_N} \frac{\tilde{a}(f) \tilde{b}^*(f)}{S_n(f)} df \right|,
  \label{eq:mismatch}
\end{align}
where the template $a$ or $b$ is taken to be complex valued containing both the
sine and cosine phases, thereby maximizing over phase, and $f_N$ is the Nyquist frequency. 
$\delta^2$ can be expressed in terms of a metric tensor $\g$ on the template
signal manifold as
\begin{align}
        \delta(\x, \dx)^2 &= \dx^T \, \g(\x)  \, \dx.
        \label{eq:mismatch_metric}
\end{align}
From the metric, we can also compute a local volume element and thereby
estimate the number of templates required to fill a given hypercube cell in the
binary tree decomposition~\cite{Owen:1995tm}:
\begin{align}
        \numper{C}(\x) &= \frac{\int \sqrt{|\det g(\x)|} dV}{ V_T },
        \label{eq:numtmps}
\end{align}
where $V_T$ is the volume of a template in mismatch space. We use the definition by Owen for the metric components in terms of the mismatch. \cite{Owen:1995tm}
\begin{equation}
  g_{ij} = - \frac{1}{2} \left[\frac{\partial^2 \delta^2(\vec{\lambda}, \Delta \vec{\lambda})}{\partial \Delta \lambda^i \partial \Delta \lambda^j}\right]_{\Delta \lambda^k=0}
  \label{eq:metric_coeffs}
\end{equation}
We have implemented two numerical schemes for estimating the metric component values that we call the iterative and deterministic methods. 
The iterative method is a standard convergence scheme for numerical differentiation leveraging the Python package numdifftools. 
The deterministic method uses definitions of the metric components as partial derivatives of the mismatch to compute the preliminary metric $\gamma_{\mu\nu}$ in a single step. 
\begin{equation}
	\begin{split}
		\gamma_{\mu\mu} &= \frac{\delta^2 (\vec{\lambda}, \Delta \vec{\lambda})}{{\Delta \lambda^\mu}^2} \\
		\gamma_{\mu\nu} &= \frac{\delta^2 (\vec{\lambda}, \Delta \vec{\lambda}) - \gamma_{\mu\mu}{\Delta \lambda^\mu}^2 - \gamma_{\nu\nu}{\Delta \lambda^\nu}^2 }{2\ \Delta \lambda^\mu\, \Delta \lambda^\nu}
	\end{split}
\end{equation}
Once the preliminary metric has been estimated using either method, we post-process the metric in two steps. First, we minimize $\gamma_{\mu\nu}\Delta\lambda^\mu\Delta\lambda^\nu$ with respect to the time lag between signals $\Delta\lambda^0$ by projecting out the time component of the metric estimate. This results in the adjusted, spatial metric components 
\begin{equation}
	g_{ij} = \gamma_{ij} - \frac{\gamma_{0i} \gamma_{0j}}{\gamma_{00}} .
\end{equation}
Where we use the term \textit{spatial} above to mean non-temporal, as in the familiar $3+1$ decomposition. 
Second, we use an eigenvalue decomposition to check for numerical stability and validity of the estimated metric. If a negative eigenvalue is found, which would incorrectly imply a negative spatial signature, we attempt a reevaluation of the metric with a coarser set of intrinsic parameters $\vec{\lambda}' = \mathrm{Coarse}(\vec{\lambda})$. \\

\newpage
\noindent The template-bank algorithm then works as follows:
\begin{enumerate}
        \item Initialize a hyper-rectangle bounding the parameter space one
          wishes to cover, e.g., a bounding box in component masses.
        \item Compute the metric $\g(\x)$ numerically at the center of the
	  hyper-rectangle. Alternatively, skip this step if the metric is
sufficiently constant.  We determine this by defining $\epsilon \equiv \left| 1
- \sqrt{|g|}_{i-2} ~/~ \sqrt{|g|}_{i-1} \right|$ and setting a threshold on
  epsilon. In other words, if the volume element of the previous two iterations
($i-2, i-1$) is sufficiently unchanged, the user may decide to skip this step.
Setting epsilon to 0 forces the metric to be recomputed.  
        \item From the metric, estimate the number of templates $\numper{C}$
needed to cover this hyper-rectangle via Eq.~\ref{eq:numtmps}.
        \item If $\numper{C}$ is greater than the user-supplied threshold
          $\numper{C}^*$, compute the side lengths of the hyper-rectangle
          according to the metric and split the cube along its largest side in
          two children cells $A$ and $B$. Call the algorithm recursively on $A$ and $B$.
        \item If $\numper{C} < \numper{C}^*$, place a template at the center of
          the cell and stop splitting. \footnote{Note that a single template is added to the bank even though $N_C$ is an estimate of the number of templates to cover a hyper-rectangle and $N_C^*$ can be greater than 1. In such a case, $N_C^*$ acts as a coarse-graining parameter. We usually set $N_C^* \leq 1$.}
\end{enumerate}
The splitting stops when all rectangles have $\numper{C} < \numper{C}^*$ or
alternatively if the user specifies a minimum coordinate volume.  In
Fig.~\ref{fig:method_illustration}, we illustrate the decomposition.

Other than waveform generation, the most computationally costly step of this
process is the evaluation of the mismatch between two templates
\eqref{eq:mismatch}, which is needed to evaluate the metric coefficients
\eqref{eq:metric_coeffs}.  In the case where the template parameter space is
bifurcated $\bifur$ times, there will be at most $2^\bifur$ hyper-rectangles.
If $\epsilon = 0$, then the metric will be evaluated for every cell and,
\begin{align}
\text{number of metric evaluations} &= \sum_{i=0}^{n} 2^i = 2^{n+1} - 1.
\end{align}
Each metric evaluation requires $\mathcal{O}(\pdim(\pdim+1)/2)$ match
calculations, where $\pdim$ is the dimension of the template parameter space,
and the exact scaling depends on the finite differencing scheme chosen. This
means that the total number of match calculations for a given bank assuming
$\epsilon = 0$ is
\begin{align}
\text{number of match evaluations} &= \frac{(2^{\bifur+1}-1) \pdim(\pdim+1)}{2} \nonumber \\
                                        &= \mathcal{O}(2^{\bifur} \pdim^2)
\end{align}
Each hyper-rectangle will contain one template, which means that a well
balanced tree will contain a bank of $\numper{B} = 2^\bifur$ templates.  Thus,
the number of match calculations \emph{per waveform} in the template bank is
\begin{align}
\frac{\text{number of match evaluations}}{\text{number of templates } (\numper{B})}
                &= \mathcal{O}(D^2).
\end{align}
The above gives a worst case scenario.  Under normal circumstances $\epsilon
>0$ and the metric is found to be sufficiently constant that it does not need
to be evaluated at the final tree depth.  This leads to typical scaling where
there are \emph{fewer match calculations than there are templates in the bank
$\numper{B}$}

By definitions the matches between waveforms used in the metric calculation are
extremely high -- approaching 1 minus floating point epsilon. Therefore, the
function of frequency is extremely smooth and we evaluate waveforms and matches
with extremely coarse spacing, typicallly 1 Hz.

\section{Results}

\begin{figure}[h!]
  \centering
  \includegraphics[width=\columnwidth]{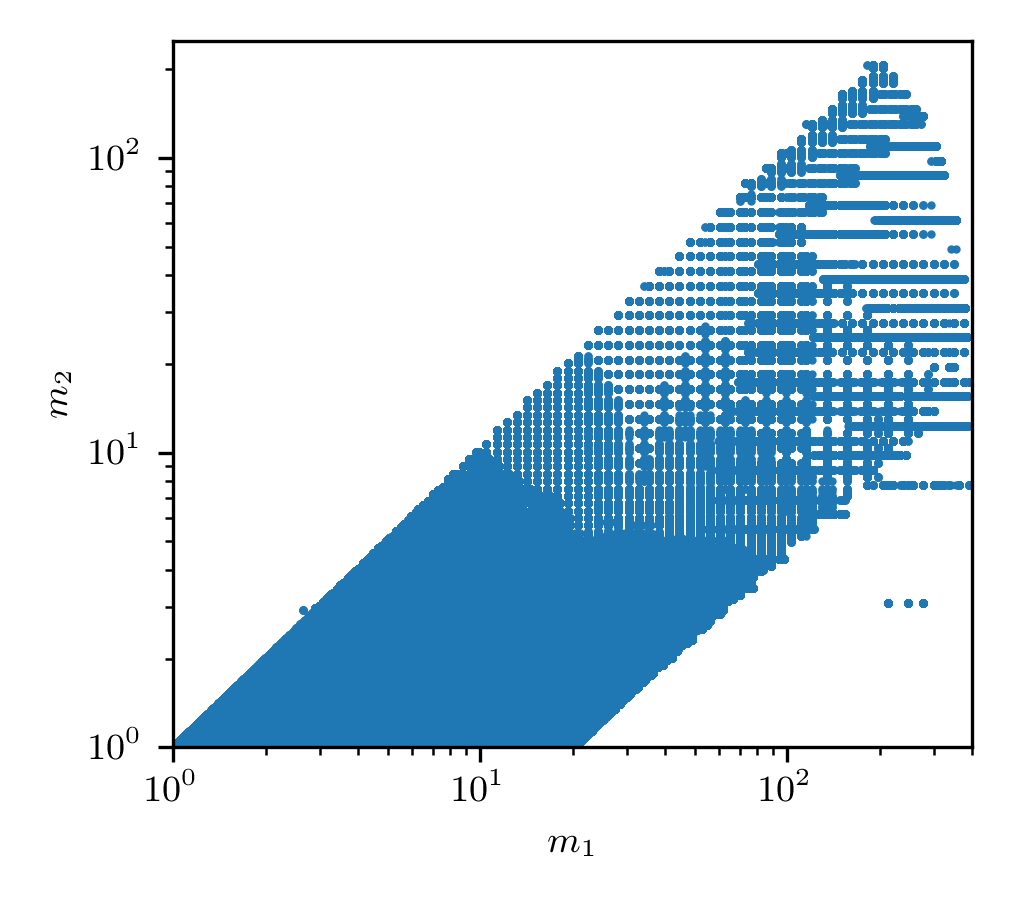}
  \caption{\label{fig:bank} 
    Example template bank. This is a projection of the three dimensional bank in coordinates $\{\log m_1, \log m_2, \chi_{\mathrm{eff}}\}$
into the $\{ \log m_1, \log m_2\}$ plane.  The templates that appear to be outside of the
region of interest have hyperrectangles that overlap with the region. Note that the naive template density is directly related to local volume element magnitude, and varies accordingly. 
    }
\end{figure}

\begin{figure*}[h!]
  \centering
  \includegraphics[width=\textwidth]{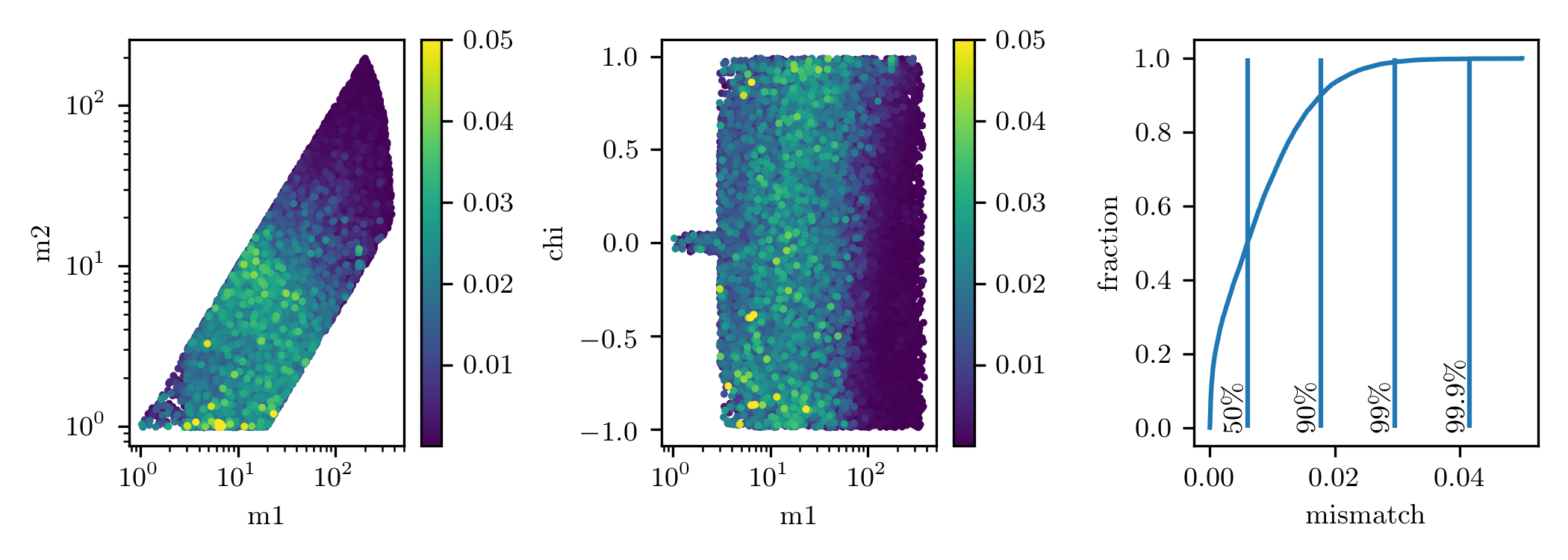}
  \caption{\label{fig:test} 
    Template bank validation. The bank achieves the requested 97\% match 99\% of
the time and a better than 98\% match 90\% of the time. The large sample
evaluation method used here is likely to be conservative since it does not
check the match of all templates in the bank.  The true performance may be
better than this. The color bar indicates mismatch of simulated signal and nearest template. The injected signals were created using uniform distributions of the individual parameters $\{\log m_1, \log m_2 \chi_{\mathrm{eff}}\}$. The bank sim maximizes match only over nearby templates because the maximum match cannot decrease by including more templates. This balances computational speed for accuracy, but preserves acceptance criteria.
    }
\end{figure*}

We used the algorithm described in the previous section to generate an advanced
LIGO template bank using projected O4 sensitivity estimates~\footnote{https://dcc.ligo.org/LIGO-T2200043}.  We
used a chirp mass range from 0.87 -- 174 M$_\odot$, a minimum secondary mass of
0.98 M$_\odot$, a maximum mass ratio of 20 and a maximum total mass of 400
M$_\odot$.  We specified an effective spin range, $\chi$, from -0.99 to 0.99
but limited the spin of objects below 3 M$_\odot$ to be less than 0.05.  We
allowed the template low frequency to go down to 10 Hz, but specified a maximum
duration of 128s.  We requested a maximum mismatch of 3\%, but also set the
minimum coordinate volume ($\Delta \log m_1 \times \Delta \log m_2 \times
\Delta \chi$) to be greater than 0.0001.  This resulted in 2,083,547 templates as
shown in Fig.~\ref{fig:bank}.

We validated the template bank by injecting 16,000 simulated signals in the
parameter space.  We find that the bank achieves the requested 97\% match
better than 99\% of the time.

\section{Conclusion}

We have described here a new method for fast template bank placement, and shown
that the method works in 3 dimensions relevant to dominant-mode
aligned-spin template searches. The \treebank ~method is computationally efficient and we
expect this method will scale to higher dimensional template placement, such as
precessing or sub-dominant mode templates, but we leave this for future work.
It should also have applications in producing high density banks for use in
rapid parameter estimation~\cite{PhysRevD.92.023002}.

A tarball containing the source code necessary to reproduce the results in this
paper can be found at \href{https://pypi.org/project/gwsci-manifold}{https://pypi.org/project/gwsci-manifold}. 

\section*{Acknowledgements}
This research has made use of data, software and/or web tools obtained from the
Gravitational Wave Open Science Center (https://www.gw-openscience.org/ ), a
service of LIGO Laboratory, the LIGO Scientific Collaboration and the Virgo
Collaboration. LIGO Laboratory and Advanced LIGO are funded by the United
States National Science Foundation (NSF) as well as the Science and Technology
Facilities Council (STFC) of the United Kingdom, the Max-Planck-Society (MPS),
and the State of Niedersachsen/Germany for support of the construction of
Advanced LIGO and construction and operation of the GEO600 detector. Additional
support for Advanced LIGO was provided by the Australian Research Council.
Virgo is funded, through the European Gravitational Observatory (EGO), by the
French Centre National de Recherche Scientifique (CNRS), the Italian Istituto
Nazionale di Fisica Nucleare (INFN) and the Dutch Nikhef, with contributions by
institutions from Belgium, Germany, Greece, Hungary, Ireland, Japan, Monaco,
Poland, Portugal, Spain.

This work was supported by National Science Foundation awards OAC-1841480,
PHY-2011865, and OAC-2103662. Computations for this research were performed on
the Pennsylvania State University's Institute for Computational and Data
Sciences gravitational-wave cluster.  CH Acknowledges generous support from the
Eberly College of Science, the Department of Physics, the Institute for
Gravitation and the Cosmos, the Institute for Computational and Data Sciences,
and the Freed Early Career Professorship.

\bibliography{references}

\end{document}